\documentclass[aps,notitlepage,a4paper,superscriptaddress,11pt]{article}

\usepackage[utf8]{inputenc}
\usepackage[dvips]{color}
\usepackage[toc]{appendix}
\usepackage[hang,flushmargin]{footmisc} 
\usepackage{titlesec,titletoc,ntheorem-hyper,authblk,abstract,latexsym,microtype,booktabs,amsmath,cleveref,fancyhdr,wrapfig,array,amsfonts, amssymb,geometry,appendix,cite}
\usepackage{dsfont}

\newcommand{\be}{\begin{equation}}
\newcommand{\ee}{\end{equation}}
\newcommand{\bea}{\begin{eqnarray}}
\newcommand{\eea}{\end{eqnarray}}
\newcommand\blfootnote[1]{%
  \begingroup
  \renewcommand\thefootnote{}\footnote{#1}%
  \addtocounter{footnote}{-1}%
  \endgroup
}

\usepackage{graphicx,lipsum}
\usepackage{caption}
\usepackage{subfigure}
\usepackage{amsmath}
\usepackage{amsfonts}
\usepackage{amssymb}
\usepackage{mathtools}
\numberwithin{equation}{section}
\usepackage{titlesec}

\numberwithin{subcase}{case}
\usepackage{framed}
\allowdisplaybreaks
\usepackage[nottoc]{tocbibind}
\usepackage[free-standing-units]{siunitx}
\usepackage{helvet}
\usepackage{url}
\usepackage{titletoc}
\usepackage{blindtext}
\usepackage[gen]{eurosym}

\title{Dirac equation with Morse potetnial under the influence of position-dependent mass and local Fermi velocity}
\author{Bijan Bagchi, Rahul Ghosh}
\affil{Physics Department, Shiv Nadar University, Gautam Buddha Nagar, \\ Uttar Pradesh 203207, India}

\begin{document}

\maketitle

\begin{abstract}
Abstract: We solve the one-dimensional Dirac equation by taking into account the possibility of position-dependence in the mass function. We also take the Fermi velocity to act as a local variable and examine the combined effects of the two on the solvability of the Dirac equation with respect to the Morse potential. Our results for the wave functions and the energy levels corresponding to such an extended scheme are furnished in closed forms.  
\end{abstract}

\blfootnote{E-mails: bbagchi123@gmail.com, rg928@snu.edu.in}

\section{Introduction}

Dirac equation is a relativistic equation that is relevant to the dynamics of spin one-half particles \cite{thal}. Solving its non-relativistic version with different variants of governing potential has been actively pursued in the literature for long. In recent times, hyperbolic graphene surface under perpendicular magnetic fields \cite{neg} has received attention not only from the perspective of the confinement of nonuniform magnetic fields of massless fermions \cite{dow1} but also from the point of view of the production of graphene crystals in the context of two-dimensional, single carbon
atom sheets (see, for instance, \cite{nov}). This has enhanced our understanding of the electronic properties of the charge carriers \cite{net, gall}. On the other hand, from a theoretical side, the techniques of supersymmetric quantum mechanics SUSYQM have been exploited by setting up first-order intertwining operators \cite{fer}. studying the influence of scalar and pseudoscalar potentials \cite{jun2020, bag2021} and looking at the general solution against spin invariant eigenstates \cite{briz}. 

In another major development of quantum mechanics, significant analysis has been carried out over the past few years on the one-dimensional position-dependent mass (PDM) problems \cite{vonroos} including adaptation of elaborate spatial structures \cite{carinena}. In a PDM scenario, an extended  Schr\"{o}dinger equation operates that depends on a generalized class of potentials possessing a set of ambiguity parameters \cite{bag2, mus, ikh}. In particular, shape invariance relation in the framework of SUSYQM has been examined and a deformed version of it within PDM was set up \cite{bag3, que}. Further, invariants were found embracing spectrum generating algebras \cite{ortiz} and consistency of indefinite effective mass was shown to exist \cite{zno}.

In the light of PDM the kinetic energy operator $\hat{T}$  acquires the form 

\begin{gather}\label{T}
 \hat{T}= -\frac{1}{4} (m^\eta(x) p m^\beta(x) p m^\gamma(x) + m^\gamma(x) p m^\beta(x) p m^\eta(x))
\end{gather}
where because of the hermiticity requirement the ambiguity parameters $\eta, \beta$ and $\gamma$ have to satisfy the constraint
\begin{equation}
    \eta+\beta+\gamma = -1
\end{equation}
Other forms of $\hat{T}$ inevitably show mutual equivalence \cite{Geller}. 

Corresponding to (1.1) the time-independent Schr\"{o}dinger equation can be cast as
\begin{gather} \label{pdmse1}
 H\phi(x)=\Big[-\frac{1}{2 } \frac{d}{dx}\frac{1}{m(x)}\frac{d}{dx}+V_{eff}(x) \Big]\phi(x)=\epsilon \phi(x)  \qquad \text{(assuming $\hbar=1$)}
 \end{gather}
 where the effective potential $V_{eff}(x)$ depends on $m(x)$ and its spatial derivatives denoted by primes
\begin{gather} \label{veffmain}
V_{eff}(x)= \mathcal{V}(x) +\frac{1}{4}(\beta+1)\frac{m''(x)}{m(x)^2}-\frac{1}{2}\left(\eta(\eta+\beta+1)+\beta+1\right)\frac{m'^2(x)}{m^3(x)}
\end{gather}
Among various types of $\hat{T}$  we focus on the Ben Daniel-Duke model involving as the choice of the ambiguity parameters, $\alpha = \gamma = 0, \beta = -1$ \cite{ben}. An advantage of it is that in such a scheme the effective potential $V_{eff}(x)$ is the same as the system potential   $\mathcal{V}(x)$.

Apart from the PDM, one can consider treating the Fermi velocity as a local variable (PDFV). Such a possibility was first proposed in \cite{dow2} to ascertain how localization effects affect graphene-like Dirac materials. Indeed, nonuniform strain in graphene suggests position-dependence could be present in the Fermi velocity. This was also revealed from scanning tunneling spectroscopy experiments \cite{juan,yan,jang}. The idea of PDFV got further impetus when gap-formation was noticed in graphene \cite{gui}. Subsequently, several interesting theoretical works have appeared where the role of PDFV was looked into \cite{mus2013, oliva, rg2022} including the study on the electronic transport in two-dimensional strained Dirac materials \cite{phan}.

The intention of the present work is to give a comprehensive solution of the Dirac equation in the presence of the Morse potential by incorporating position-dependence of the mass as well taking into consideration the local behaviour of the Fermi velocity. As is well known Morse potential occupies an important place in quantum mechanics in that is of much use in spectroscopic applications \cite{pau}. Moreover, it could be connected with the Coulomb potential under a coordinate transformation \cite{hay, lah, bag4}.

The paper is organised as follows: 

The next section summarizes the basic ingredients of SUSYQM. In section 3 we write the Dirac equation in full in the presence of PDM and PDFV and describe the mathematical formulation our model in terms of a pair of coupled differential equations for the two-component wavefunctions. In section 4, we give the complete solution of the Morse problem. Finally, in section 5, a summary of our work is presented. 
 
 \newpage
 
 \section{Aspects of SUSYQM}
 
 The basic formalism \cite{jun1996, bag2000, kha2001, iof2006, gan2017} of SUSYQM involves the supercharges $Q, Q^\dagger$ that depict fermionic-like properties. The governing Hamiltonian can be expressed in the manner

\begin{equation}
    \mathcal{H} = \{Q, Q^\dagger \}
\end{equation}
The basic properties of $Q, Q^\dagger$ are 

\begin{eqnarray}
&&    (Q)^2 = 0 = (Q^\dagger)^2 \\
&& [Q, \mathcal{H}] = 0 = [Q^\dagger, \mathcal{H}]
\end{eqnarray}
To represent them, we introduce the corresponding operators $O$ and $O^\dagger$ and write

\begin{equation}
Q = \mathcal{O} \otimes \sigma_-, \quad Q^\dagger = \quad \mathcal{O}^\dagger \otimes \sigma_+
\end{equation}
where the quantities $\sigma_{\pm}$ denote $\sigma_{\pm} = \frac{1}{2}(\sigma_1 \pm i \sigma_2)$, $\sigma_1$ and $\sigma_2$ are the usual Pauli matrices.  Taking a first-order differential realization of $\mathcal{O},\mathcal{O}^\dagger$  

\begin{equation}
    \mathcal{O} = \frac{\hbar}{\sqrt{2m}}\partial + \mathcal{W}(x), \quad \mathcal{O}^\dagger = -\frac{\hbar}{\sqrt{2m}}\partial + \mathcal{W}(x)
\end{equation}
where $m$ is the mass, and $\mathcal{W}(x)$ is the superpotential of the system, we can project $Q$ and $Q^\dagger$ in the matrix forms

\begin{equation}
   Q = \left( \begin{array}{cc} 0 & 0  \\ \frac{\hbar}{\sqrt{2m}}\partial + \mathcal{W}(x)  & 0  \end{array} \right), \quad Q^\dagger =  \left( \begin{array}{cc} 0 &  -\frac{\hbar}{\sqrt{2m}}\partial + \mathcal{W}(x) \\0 & 0  \end{array} \right)
\end{equation}
where $\partial \equiv \frac{d}{dx}$. The Hamiltonian $\mathcal{H}$ is thus rendered diagonal 

\begin{equation}
   \mathcal{H} = \left( \begin{array}{cc} \mathcal{H}_+ & 0  \\  0 & \mathcal{H}_- \end{array} \right)
\end{equation}
with the components $H_{\pm}$ given by the factorization \cite{miel2004}
\begin{gather}
 \mathcal{H}_+ = \mathcal{O}^\dagger \mathcal{O}= -\frac{\hbar^2}{2m}\partial^2 + U^{(+)} (x) - \Lambda \\
 \mathcal{H}_- = \mathcal{O} \mathcal{O}^\dagger  = -\frac{\hbar^2}{2m}\partial^2 + U^{(-)} (x) - \Lambda   
\end{gather}
These are in typical Sch\"{r}odinger form defined at some cut-off energy value $\Lambda$. 

The SUSY partner potentials $U^{(+)}$ and $U^{(-)}$ can be projected as

\begin{equation}{\label{V+-}}
    U_{\pm} (x) = \mathcal{W}^2(x) \mp \frac{\hbar}{\sqrt{2m}} \mathcal{W}'(x) + \Lambda
\end{equation}
where the prime represents a derivative with respect to $x$. For unbroken $SUSY$ the ground state wavefunction $\psi_0^+ (x)$ is non-degenerate which we can associate with the component $\mathcal{H}_+$. This has the implication

\begin{eqnarray}
&& \mathcal{O} \psi_0^+ (x) = 0 \\
&& \implies \psi_0^+ (x)  \propto \exp \left ( -\frac{\sqrt{2m}}{\hbar}\int^x \mathcal{W}(t) dt \right )
\end{eqnarray}
reflecting that the normalizability of $\psi_0^+ (x)$ restricts the superpotential to obey $\int^x W(t) dt > 0$ as $x \rightarrow \infty$. 

The double-degeneracy of the spectrum is guided by the following
intertwining relationships of $\mathcal{H}$

\begin{equation}
    \mathcal{O} \mathcal{H}_+ = \mathcal{H}_- \mathcal{O}, \quad  \mathcal{H}_+ \mathcal{O}^\dagger= \mathcal{O}^\dagger\mathcal{H}_- 
\end{equation}
and furnishes the isospectral connections between $\mathcal{H}^{(+)}$ and $\mathcal{H}^{(-)}$. These are of course consistent with the projections $(2.8)$ and $(2.9)$.    

\section{PDM and PDFV}

In the standard constant mass $m_0$ and constant Fermi velocity $v_F$ case the one-dimensional Dirac Hamiltonian is given by \cite{jun2020} (see also \cite{alh2011})

\begin{equation}\label{H_D1}
  H_D = v_f \sigma_x p_x  +\sigma_y W(x)  +\sigma_z  m_0 v_f^2 +\mathds{1} V(x) 
\end{equation}
where $V(x)$ is the electrostatic potential, $W(x)$ is the pseudoscalar potential and $\mathds{1}$ is the block-diagonal unit matrix. In \cite{bag2021}, $W(x)$ was interpreted to behave like the superpotential of the system. The Pauli matrices are known to be
\begin{gather}
\sigma_x = \left( \begin{array}{cc} 0 & 1  \\ 1 & 0  \end{array} \right), \quad \sigma_y = \left( \begin{array}{cc} 0 & -i  \\ i & 0  \end{array} \right), \quad \sigma_z = \left( \begin{array}{cc} 1 & 0  \\ 0 & -1  \end{array} \right) 
\end{gather}

The material properties of the Dirac particle are carried by the combined effects of the Fermi velocity and the mass function that produce a heterostructure \cite{mele}. The possibility of the Fermi velocity varying from material to material is quite comprehensible in condensed matter physics \cite{per2009}. This motivates us to look into a situation when both the mass function and Fermi velocity become local function of position. In setting up of our scheme we ignore the effects of the electrostatic potential $V$ as is suggested by the analysis of the intertwining relations \cite{jun2020, ish}.

  We therefore consider PDM as well as PDFV to rewrite (\ref{H_D1}) as follows 
\begin{gather} \label{H_D}
  H_D = \sqrt{v_f(x)} \sigma_x p_x \sqrt{v_f(x)} +\sigma_y W(x)  +\sigma_z  m(x) v_f^2 (x)
\end{gather}
In the two-dimensional matrix form $H_D$ becomes

\begin{gather}\label{H_Dmatrix}
     H_D = \left( \begin{array}{cc} m v_f^2  &  -i\hbar \sqrt{v_f} \partial \sqrt{v_f} -iW  \\ -i\hbar \sqrt{v_f} \partial \sqrt{v_f}  + i W & - m v_f^2   \end{array} \right)
\end{gather}
where note that $m = m(x)$ and $v_f = v_f (x)$. Further, $V$ and $W$ are arbitrary function of $x$. 

Operating on a spinor with components $ (\psi_+ \quad \psi_-)^T$ emerges the coupled pair of equations  

 \begin{gather}
 (-i\hbar \sqrt{v_f} \partial \sqrt{v_f} -i W) \psi_- = D_- \psi_+ \end{gather}
 \begin{gather} \label{lowercomponent}
 (-i\hbar \sqrt{v_f} \partial \sqrt{v_f} +i W) \psi_+ = D_+ \psi_- 
\end{gather}
where $E$ is the energy eigenvalue and we have chosen to work with natural units $\hbar = 1$. In the above the quantities $D_{\pm}$ stand for $D_{\pm} = E \pm m v_f^2$. 
Specifically the equation for the upper component reads
\begin{gather}
-\frac{v_f^2}{D_+}\frac{d^2 \psi_+}{dx^2} - \frac{d}{dx}\Big(\frac{v_f^2}{D_+}\Big) \frac{d\psi_+}{dx} + \Big[\frac{1}{D_+} \Big(W^2 -\frac{1}{4}{v'_f}^2-\frac{1}{2}v_f v''_f\Big) + v_f\frac{d}{dx} \Big(\frac{W}{D_+}\Big)     \nonumber \\ 
- \frac{1}{2}v_f v'_f \frac{d}{dx}\Big(\frac{1}{D_+}\Big)  \Big] \psi_+ = D_- \psi_+
\end{gather}
while for the lower component we have

\begin{gather}
-\frac{v_f^2}{D_-}\frac{d^2 \psi_-}{dx^2} - \frac{d}{dx}\Big(\frac{v_f^2}{D_-}\Big) \frac{d\psi_-}{dx} + \Big[\frac{1}{D_-} \Big(W^2 -\frac{1}{4}{v'_f}^2-\frac{1}{2}v_f v''_f\Big) -v_f\frac{d}{dx}\Big(\frac{W}{D_-}\Big)     \nonumber \\ 
- \frac{1}{2}v_f v'_f \frac{d}{dx}\Big(\frac{1}{D_-}\Big)  \Big] \psi_- = D_+ \psi_-
\end{gather}

To proceed further we employ the constancy condition \cite{mus2013} 

\begin{equation}
    m(x)v_f^2(x)= \mbox{constant} = m_0 v_0^2
\end{equation}
where $m_0$ and $v_0$ are both constants. This helps greatly in the reduction of (3.7) and (3.8). In particular both $D_{\pm}$ are rendered constant. As a result the coupled equations (3.7) and (3.8) are disentangled leaving us with two separate equations, one for $\psi_+$ and another for $\psi_-$
\begin{gather} \label{upperDirac}
- v_f^2 \frac{d^2 \psi_+}{dx^2} - \frac{d}{dx}(v_f^2) \frac{d\psi_+}{dx} + \Big[ \Big(W^2 -\frac{1}{4}{v'_f}^2-\frac{1}{2}v_f v''_f\Big) + v_f W' )\Big] \psi_+ = (E^2-m_0^2v_0^4) \psi_+     
\end{gather}
\begin{gather}\label{lowerDirac}
- v_f^2 \frac{d^2 \psi_-}{dx^2} - \frac{d}{dx}(v_f^2) \frac{d\psi_-}{dx} + \Big[ \Big(W^2 -\frac{1}{4}{v'_f}^2-\frac{1}{2}v_f v''_f\Big) - v_f W' )\Big] \psi_- = (E^2-m_0^2v_0^4) \psi_-
\end{gather}

Making a change of variable through defining
\begin{gather}
    \psi_\pm(x) = \frac{1}{\sqrt{v_f(x)}} \Phi_\pm (y(x))  
\end{gather}
where the new quantity $y(x)$ stands for the integral
\begin{equation}
     y(x)= \int^x \frac{dz}{v_f(z)}+ \mbox{constant}
\end{equation}
results in the set of differential equations
\begin{gather}\label{newSE}
    - \frac{d^2 \Phi_\pm(y)}{dy^2}  + \Big[ W^2(x(y)) \pm v_f(x(y)) W'(x(y)) \Big] \Phi_\pm(y) =  (E^2- m_0^2 v_0^4) \Phi_\pm(y)
\end{gather}
We thus have the SUSY form for the partner potentials as defined below albeit in terms of the variable $y$ 
\begin{gather}\label{extendedV+-}
    V_\pm(x(y))=W^2(x(y)) \mp v_f(x(y)) W'(x(y))
\end{gather}
These match with $U_{\pm}$ in $(2.10)$ for $\Lambda = 0$.

An interesting off-soot is that if we compare the basic PDM guided Sch\"{o}dinger equation (1.3) with either (3.10) or (3.11) the mass function $m(x)$ coincides with the inverse square of the Fermi velocity \cite{rg2022}

\begin{gather}\label{mass}
 m(x)= \frac{1}{2v_f^2(x)}
\end{gather}
This result has a significance in the context of deformed shape-invariance within SUSYQM. In fact, if we look at the equation for the positive-definite deforming function $f$ introduced in \cite{bag3} namely

\begin{equation}
    H \phi (x) = \left [- \left ( \sqrt{f(x)} \frac{d}{dx} \sqrt{f(x)} \right )^2 + V_{eff} (x) \right ] \phi (x) = E \phi (x)
\end{equation}
(which follows as a consequence of (1.3) by implementing a deformed shape-invariance relation), then such a function $f$ is easily recognized to play the role of PDFV. 

With this background we now turn to the Morse potential.

\section{Morse potential}

We assume for $W(x)$ the following choice

\begin{gather}
   W (x)= 
   \begin{cases}
     \omega_0-\omega_1 x \quad \text{$x > 0$}  \\
     0  \quad \text{$x \leq 0$} 
\end{cases}
\end{gather}
where $\omega_0$ and $\omega_1$ are positive constants, and take for the PDFV the representation
\begin{gather}
v_f (x)=
\begin{cases}
     \alpha x \quad \text{$x > 0$}  \\
     0  \quad \text{$x \leq 0$} 
\end{cases}
\end{gather}
where $\alpha$ is a positive constant. From (3.10) the mass function $(\ref{mass})$ turns out to be
\begin{gather}
   m(x)=  \frac{1}{2\alpha^2 x^2} \quad \text{$x > 0$}
\end{gather}
As a result, from $(\ref{extendedV+-})$, the extended SUSY partner potentials read
\begin{gather}
   V_\pm(x)= 
   \begin{cases}
     \omega_0^2+\omega^2_1 x^2 -2\omega_1(\omega_0\pm \frac{\alpha}{2}) x \quad \text{ for $x > 0$}  \\
     0  \quad \text{ for $x \leq 0$} 
   \end{cases}
\end{gather}
which correspond to shifted harmonic oscillator potentials. 

If we now make a change of variable to  $t=\frac{1}{\alpha} \ln x \in (-\infty,\infty)$, the partner potentials get transformed to the forms
\begin{gather}
   V_\pm(t)= \omega_0^2+\omega^2_1 e^{2\alpha t} -2\omega_1(\omega_0\pm \frac{\alpha}{2}) e^{\alpha t}, \quad t \in (-\infty,\infty)
\end{gather}
Clearly, the two cases in (4.5) correspond to the potentials of the one-dimensional Morse potential.

The accompanying Schr\"{o}dinger equation in the $t$-variable for $\Phi_+(t)$ reads
\begin{gather} \label{SEmorse}
     \frac{d^2 \Phi_+(t)}{dt^2}  + \left [k^2-\omega_0^2+\omega^2_1 e^{2\alpha t} -2\omega_1(\omega_0 - \frac{\alpha}{2}) e^{\alpha t} \right] \Phi_+(t) =  0
\end{gather}
where $k^2=E^2-\frac{1}{4}$. Its solution in terms of associated Laguerre polynomials $L_n^\kappa (\xi)$ is \cite{lan1991,kha2001}
\begin{gather}
   \Phi_+ (\xi) \propto \xi^{\frac{\kappa}{2}} e^{-\frac{\xi}{2}} L_n^\kappa \left( \xi \right)  
\end{gather}
where $\kappa = \frac{2\omega_0}{\alpha}-2n \in \Re$ and $\xi= \frac{2\omega_1 }{\alpha} e^{\alpha t}$ $\in (0,\infty)$, along with the eigenvalues 
\begin{equation}
E_n^2= \frac{1}{4}+\omega_0^2-(\omega_0 -\alpha n)^2 
\end{equation}
The allowed values for $n$ are $n = 0, 1, . . . , n_{max} (n_{max} <\frac{\omega_0}{\alpha})$.

Reverting to $x$-coordinate, $(4.7)$ moves over to the representation
\begin{gather}
  \psi_+(x) = N e^{-\frac{\omega_1 x}{\alpha} } x^{\frac{\kappa-1}{2}} L_n^\kappa \left( \frac{2\omega_1 x}{\alpha} \right) 
\end{gather}
where $N$ is the normalization constant. As a check, note that if we take the Fermi velocity to be constant scaled to unity, $\psi_+(x)$ will be same as furnished in \cite{bag2006}. For completeness we also provide the solution for $\psi_{-}(x)$ 

\begin{flalign}
\psi_{-}(x) =\frac{N i e^{-\frac{\omega_1 x }{\alpha }} x^{\frac{\kappa-2}{2}}}{2 \sqrt{
\alpha } \mathcal{D}} 
& \Bigg[4 \omega_1 x  L_{n-1}^{\kappa+1}\left(\frac{2  \omega_1 x}{\alpha }\right)   \notag \\    
&  +(\alpha + 2 n \alpha + 4  \omega_1 x) L^{\kappa}_n \left(\frac{2 \omega_1 x}{\alpha }\right) \Bigg]
\end{flalign}
where the quantity $\mathcal{D}$ is given by  $\mathcal{D} =\sqrt{\frac{1}{4}+\omega_0^2-(\omega_0 -\alpha n)^2}+\frac{1}{4}$.
(4.9) and (4.10) give the complete closed-form solutions of the Morse partner wavefunctions in the presence of PDM and PDFV along with energy levels (4.8).

\section{Summary}

To summarize, we have shown in this paper that the Dirac equation in the presence of the PDM and PDFV admits of an exact solution for the Morse potential. Employing the factorization method and using von Roos prescription of the modified kinetic energy term for the PDM, and further observing that the underlying pseudoscalar potential acts as the superpotential of the system in the setup of SUSYQM, we solved a pair of coupled equations containing the components of the Dirac spinor. In this regard we utilized the result that the mass function behaves like the inverse square of the Fermi velocity. Through disentanglement we derived the complete solution of the Morse-like partner potentials.

\section{Acknowledgment}

One of us (RG) thanks Shiv Nadar University for the grant of senior research fellowship.

\section{Data availability statement}

All data supporting the findings of this study are included in the article.

\end{document}